\pdfoutput=1
\documentclass[12pt]{article}
\usepackage{epsfig,amssymb,amsmath}

\def\Title#1{\begin{center} {\Large {\bf #1} } \end{center}}

\begin{document}

\begin{center}
\sl Proceedings of CKM 2012, the 7th International Workshop on the CKM Unitarity Triangle, University of Cincinnati, USA, 28 September - 2 October 2012
\end{center}
\vspace{1cm}
\Title{Top Decays in the Standard Model and Beyond}

\bigskip\bigskip


\begin{raggedright}  

{\it Jure Drobnak\index{Drobnak, J.}\\
Jozef Stefan Institute \\
Jadranska 19,\\
1000 Ljubljana, SLOVENIA}
\bigskip\bigskip
\end{raggedright}

\section{Introduction}

Top quark is by far the heaviest fermion in Standard Model (SM) and as such interesting when it comes to searches for physics beyond the SM. Precision studies of top quark properties are currently underway. In these proceedings we explore the possibility of new physics (NP) manifesting itself in rare top quark decays. In particular, we adopt an effective theory description of FCNC top quark decays and possible deviations from the SM form of $tWb$ vertices.

FCNC top quark decays are being searched for at LHC and Tevatron. Due to the smallness of branching fractions predicted by SM, such decays are considered unobservable, meaning that potential experimental detection would indicate presence of NP. On the other hand, NP in charged quark currents impacting $tWb$ vertices could be observed through the measurement of the helicity fractions of $W$ bosons produced in the main decay channel of the top quark $t\to b W$.

Due to the outstanding role of the top quark in rare processes of meson physics, one should also consider the effects of NP in top quark sector on the well measured and theoretically understood processes in $B$ and $K$ physics.

\section{NP in top quark FCNC decays}
Within SM $t \to q V$ decays, where $q=u,c$ and $V = Z,\gamma,g$ are highly suppressed~\cite{AguilarSaavedra:2004wm}, with branching fractions of $\mathrm{Br}[t\to c V]\sim 10^{-14}-10^{-12}$, way below the reach of experiments. Various models of NP can lift this suppression \cite{Yang:2008sb}. We parametrize NP manifestation in form of FCNC effective vertices as
\begin{eqnarray}
{\mathcal L}_{\mathrm{eff}} = \frac{v^2}{\Lambda^2}a_L^{Z}\mathcal O_{L}^Z
+\frac{v}{\Lambda^2}\Big[b^{Z}_{LR}\mathcal O_{LR}^{Z}+b^{\gamma}_{LR}\mathcal O_{LR}^{\gamma}+b^{g}_{LR}\mathcal O_{LR}^{g}
\Big] + (L \leftrightarrow R) + \mathrm{h.c.}\,. \label{eq:Lag}
\end{eqnarray}
For the definition of operators see Ref.~\cite{Drobnak:2010wh}. Turning first to the indirect constraints of NP from meson physics, a detailed study has been performed in Ref.~\cite{Fox:2007in} and has shown that apart from $\mathcal O_L$ operator, indirect constraints are not stringent enough to forbid the direct observation of FCNC decays at LHC\footnote{Matching between $SU(2)_L$ invariant operators used in Ref.~\cite{Fox:2007in} and those used in Eq.~(\ref{eq:Lag}), can be found in Ref.~\cite{Drobnak:2008br}}.

Since there has, as of now, been no observation of FCNC top quark processes, upper 95\% C.L. bounds on branching fractions have been obtained. In particular, ATLAS has set the following limit through single top production~\cite{Aad:2012gd}
\begin{eqnarray}
\mathrm{Br}[t\to u,c g] < 5.7\times 10^{-5}\,, 2.7\times 10^{-4}.
\end{eqnarray}
In addition, ATLAS~\cite{Aad:2012ij} and CMS~\cite{:2012sd} report the following limits for the $t\to q Z$ decays
\begin{eqnarray}
\mathrm{Br}[t\to q Z ] < 7.3\times 10^{-3}\,, 2.4\times 10^{-3}\,,
\end{eqnarray}
respectively. The bounds for photonic modes come from HERA~\cite{Aaron:2009vv} and CDF~\cite{Abe:1997fz}
\begin{eqnarray}
\mathrm{Br}[t\to u \gamma] < 6.4\times 10^{-3}\,,\hspace{0.5cm}\mathrm{Br}[t\to q \gamma] = 3.2 \times 10^{-2}\,,
\end{eqnarray}
respectively. What is more, the study of projected sensitivities of ATLAS to $Z$ and $\gamma$ decay modes~\cite{Carvalho:2007yi} predict probing branching fractions of the order $\nolinebreak{\mathrm{Br}[t\to q\gamma,Z] \sim 10^{-5}}$ with 100 $\mathrm{fb}^{-1}$ of collected data, promising substantial improvements of bounds in case no FCNC processes are observed.

\begin{figure}
\begin{center}
\includegraphics[width=0.45\textwidth]{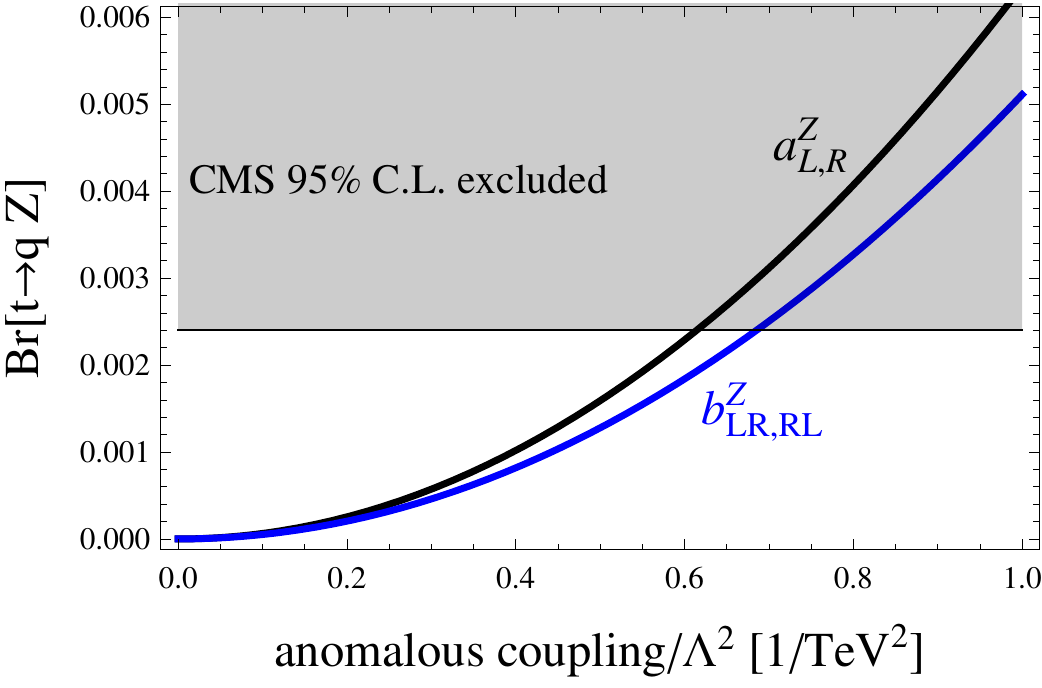}\hspace{0.05\textwidth}
\includegraphics[width=0.45\textwidth]{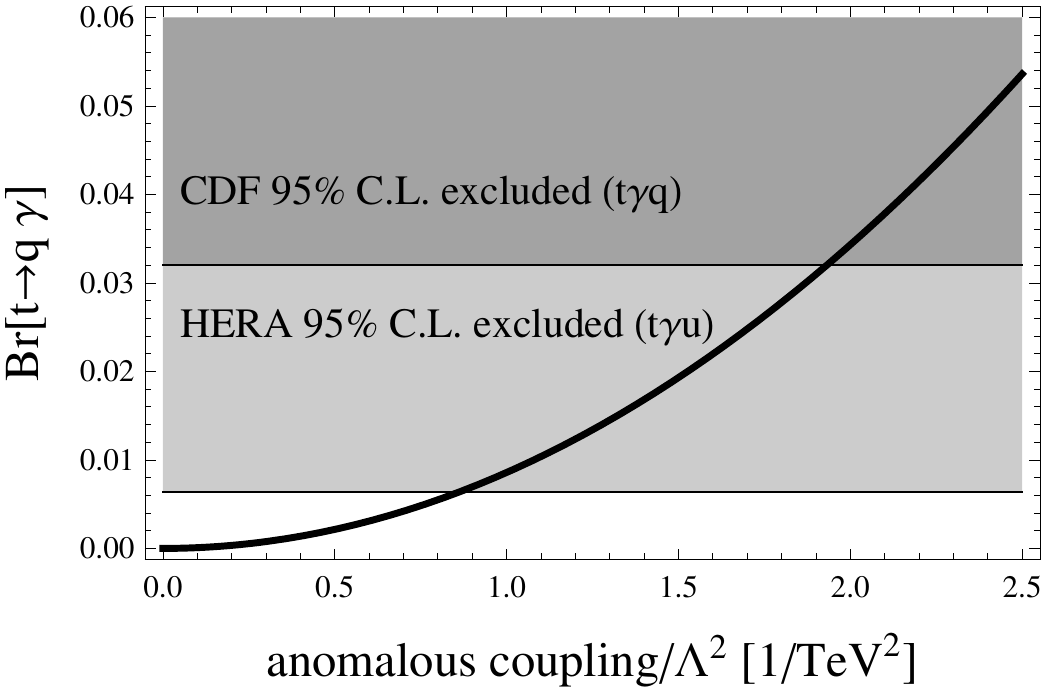}

\caption{Graphs depicting the extraction of 95\% C.L. bounds on anomalous top quark FCNC couplings through $t\to q Z$ (left) and $t\to q \gamma$ (right) decays.}
\label{fig:1}
\end{center}
\end{figure}

In turn these bounds can be used together with theoretical predictions given in Refs.~\cite{Drobnak:2010wh, Zhang:2008yn} to constrain the effective couplings characterizing NP. Figure~\ref{fig:1} summarizes the analysis for $Z$ and $\gamma$ channels.
The obtained 95\% C.L. bounds for all three channels read
\begin{eqnarray}
\begin{array}{cl|cl}
b^\gamma/\Lambda^2 < 0.86 \,\, \mathrm{TeV}^{-2}\,, & \mathrm{for}\,\, q = u\,, 
&b^g/\Lambda < 6.9 \times 10^{-3}\,\, \mathrm{TeV}^{-1}\,,& \mathrm{for}\,\, q=u\,,\\
b^\gamma/\Lambda^2 < 1.93 \,\, \mathrm{TeV}^{-2}\,, & \mathrm{for}\,\, q = u,c\,,
&b^g/\Lambda < 1.6 \times 10^{-2}\,\, \mathrm{TeV}^{-1}\,,&\mathrm{for}\,\, q=c\,, \\\hline
a^Z/\Lambda^2  < 0.62\,\, \mathrm{TeV}^{-2}\,, & \mathrm{for}\,\, q = u,c\,,\\
b^Z/\Lambda^2  < 0.69\,\, \mathrm{TeV}^{-2}\,, & \mathrm{for}\,\, q = u,c\,,
\end{array}
\end{eqnarray}
where the results for the gluonic anomalous couplings are derived in Ref.~\cite{Aad:2012gd}.


\section{NP in charged quark currents}
Similarly to previous section, we can parametrize the effects of new physics in $tWb$ vertex in term of anomalous couplings 
\begin{eqnarray}
\mathcal L_{\mathrm{eff}} = a_L \mathcal O_L + b_{LR} \mathcal O_{LR} + (L\leftrightarrow R) + \mathrm{h.c.}\,,
\end{eqnarray} 
where $\mathcal O_L$ and $\mathcal O_{LR}$ operators are of vector and dipole type respectively and contain $t$, $b$ and $W$ fields\footnote{For exact definition see Ref.~\cite{Drobnak:2010ej}.}.

Detailed analysis of indirect constraints from $\Delta B =1,2$ processes and electro-weak precision observables has been performed in Refs.~\cite{Drobnak:2011aa, Grzadkowski:2008mf, Greiner:2011tt, Zhang:2012cd}. The upshot is that Wilson coefficients accompanying operators $\mathcal O_R$ and $\mathcal O_{RL}$ and evaluated at the electroweak scale, are severely constrained $|a_R,b_{RL}|\lesssim 0.001 $(mostly through $b\to s \gamma$), while bounds for other Wilson coefficients are above percent level.

The structure of $tWb$ vertex can be probed at LHC and Tevatron through the analysis of helicity fractions $\mathcal F_i$ of $W$ bosons produced in the main decay channel. Performing a naive average of Tevatron~\cite{Aaltonen:2012rz} and ATLAS~\cite{Aad:2012ky} results we obtain the following experimental values to be compared with the state of the art SM predictions~\cite{Czarnecki:2010gb}
\begin{eqnarray}
\begin{array}{lclclc}
\mathcal F_L &=& 0.692\pm 0.053\,,&
\mathcal F_+ &=&-0.013\pm 0.034\,,\\
\mathcal F_L^{\mathrm{SM}} &=& 0.687 \pm 0.005 \,, & \mathcal F_+^{\mathrm{SM}} &=& 0.0017 \pm 0.0001\,.
\end{array}
\end{eqnarray}
The small value of SM prediction for the chiraly suppressed $\mathcal F_+$ makes this observable very appealing for NP searches, since a measured value of percent level would clearly indicate NP governing the decay.

\begin{figure}[h!]
\begin{center}
\includegraphics[width=0.45\textwidth]{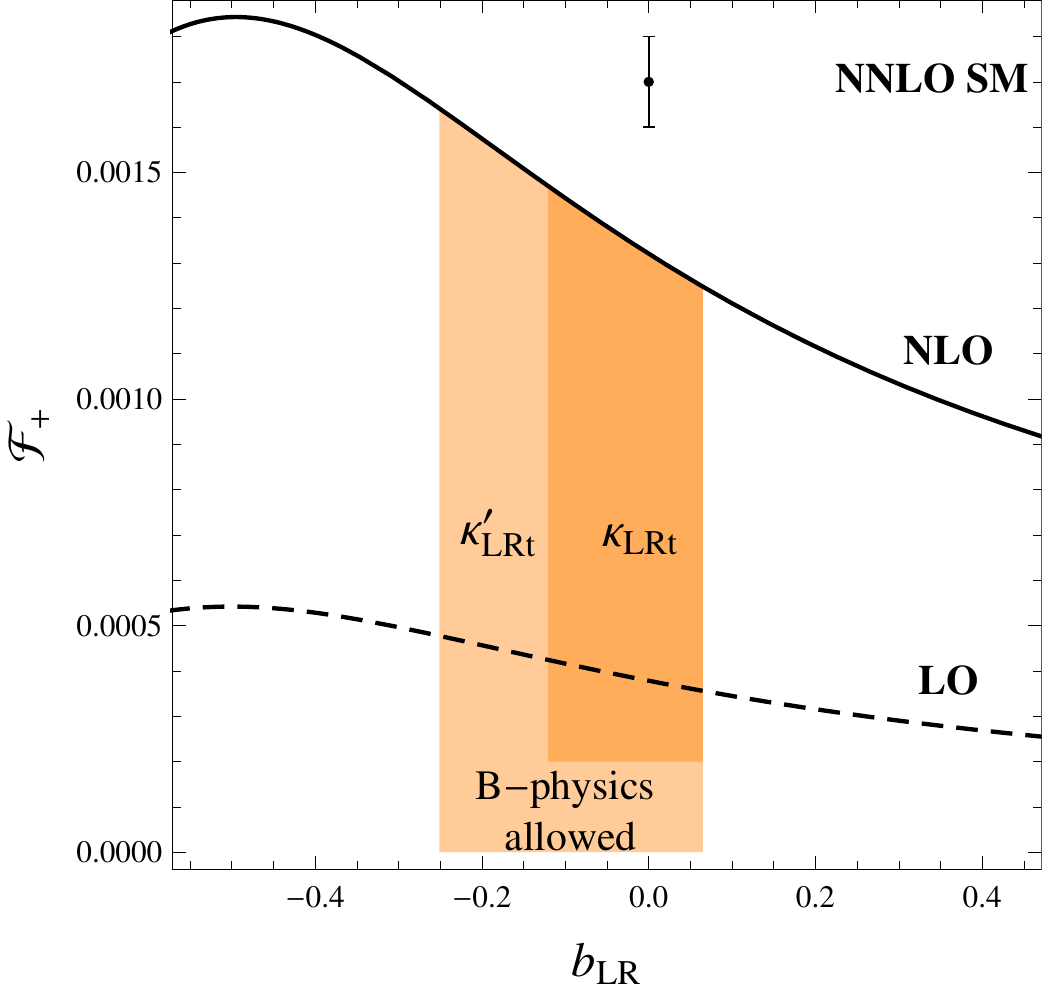}\hspace{0.05\textwidth}
\includegraphics[width=0.45\textwidth]{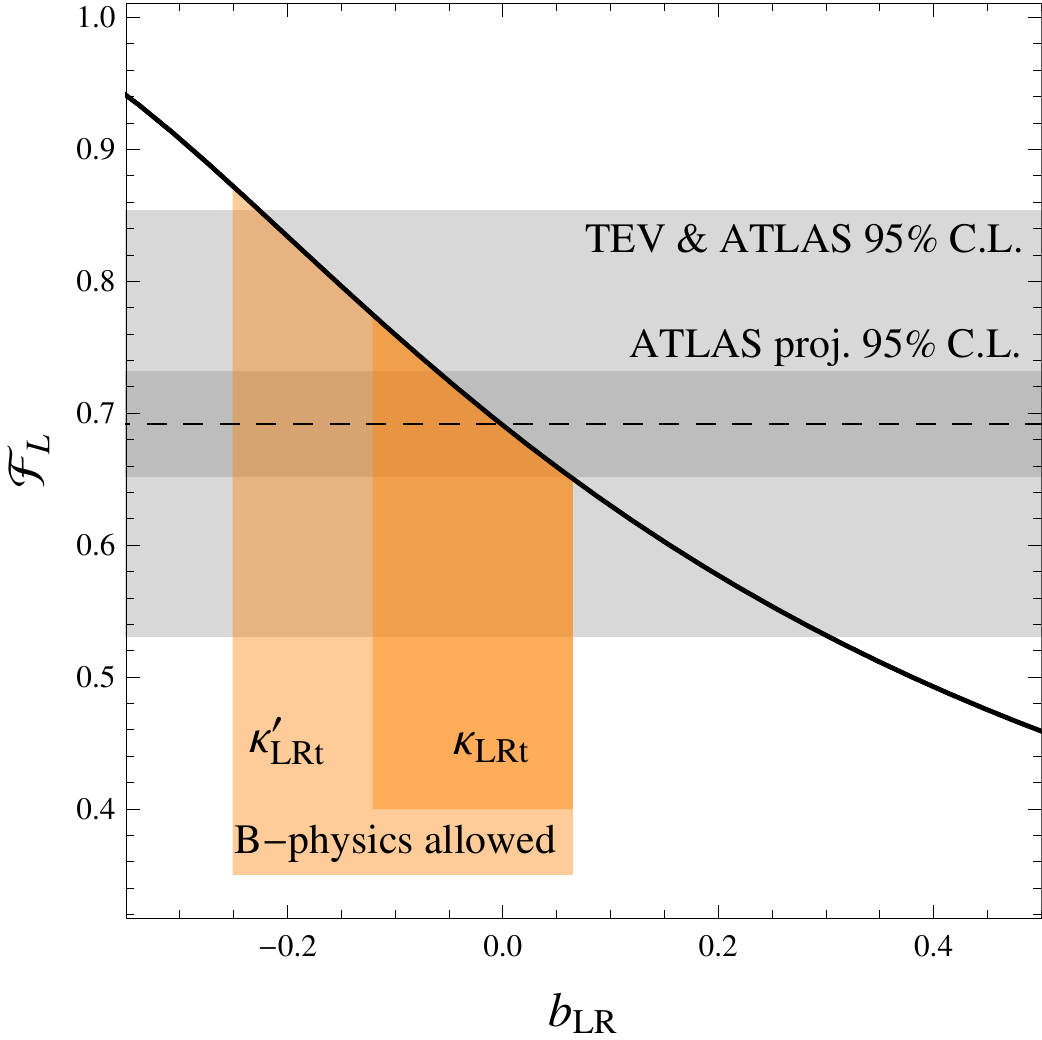}
\caption{Dependence of $\mathcal F_+$ (left) and $\mathcal{F}_L$ (right) on anomalous coupling $b_{LR}$ which is considered to be real. Orange bands represent 95\% C.L. intervals obtained from indirect considerations. On the right graph we show the combined experimental central value (dashed) and the $95\%$ C.L. band along with the projected ATLAS band. }
\label{fig:2}
\end{center}
\end{figure}

In Figure~\ref{fig:2} we explore the effects of anomalous coupling $b_{LR}$ on the helicity fraction. We only consider this coupling since $a_R, b_{RL}$ are constrained from indirect considerations to such extent that we cannot expect them to notably impact the helicity fractions. On the other hand, $a_L$ cancels out of the helicity fraction expressions as long as only one anomalous coupling is considered to be non-zero at a time.
From the graph on the lefthand side we can see that variation of the coupling $b_{LR}$ can not, even at next-to-leading order in QCD, increase the helicity fraction to percent order or higher. Furthermore, from the righthand side graph we can infer that for the $b_{LR}$ coupling the direct constraints from helicity fraction $\mathcal{F}_L$ are competitive with the indirect and are further expected to constrain the coupling in the future. 

\section{Conclusions}
In conclusion, rare top quark decays are interesting for NP considerations. Presence of NP in presented observables could easily be observed, however the SM compatible measurements are also beneficial since they serve to constrain various models of NP. Some of these direct constraints are for the first time becoming competitive with the indirect constraints from meson physics.

\end{document}